\documentstyle[12pt]{article}
\begin{document}
\begin{titlepage}
\hspace*{9.5cm}{\bf HU-SEFT R 1993-15}
\vskip 1.0 cm
\begin{center}
\vspace*{1.0cm}
{\Large Quantum Poincar\'e Subgroup of q-Conformal Group \\
and q-Minkowski Geometry }
\vskip 1.0cm
by
\vskip 0.5 cm

{\large M. Chaichian}\renewcommand{\thefootnote}{\ }\footnote{To appear in the
Proceedings of the Workshop "Generalized Symmetries in Physics", (July 1993,
Clausthal), editors: H.D. Doebner et al., to be published by
World Scientific.}\renewcommand{\thefootnote}{*}
\footnote{High Energy Physics
Laboratory, Department of Physics, P.O.Box 9 (Siltavuorenpenger 20 C),\\
SF-00014, University of Helsinki, Finland\\\hspace*{6.5cm} and\\ Research
Institute for
High Energy Physics, P.O.Box 9 (Siltavuorenpenger 20 C), SF-00014, University
of Helsinki, Finland;
E-mail address: CHAICHIAN@phcu.Helsinki.Fi} and
{\large A.P.Demichev}\renewcommand{\thefootnote}{\dagger}\footnote{Nuclear
Physics Institute, Moscow State University, 119899, Moscow, Russia; E-mail
address: DEMICHEV@COMPNET.MSU.SU}
\end{center}
\vspace{3 cm}
\begin{abstract}
\normalsize
We construct quantum deformation of Poincar\'e group using as a
starting point $SU(2,2)$ conformal group
and twistor-like definition of the Minkowski space. We
obtain quantum deformation of $SU(2,2)$ as a real form of multiparametric
$GL(4,C)_{q_{ij},r}$. It is shown that Poincar\'e subgroup exists for
special nonstandard one-parametric deformation only, the
deformation parameter $r$ being equal to unity. This leads to
commuting affine structure of the corresponding Minkowski space
and simple structure of the corresponding Lie algebra, the
deformation of the group being non-trivial.

\end{abstract}
\end{titlepage}

\section{Introduction}

At present there is well developed theory of quantum semisimple
matrix Lie groups and algebras \cite{FRT}- \cite{Jim}. But as is well
known in classical case, Minkowski geometry is intrinsically
connected with inhomogeneous Poincar\'e group. General theory of
q-deformation of inhomogeneous groups is absent. Method of quantum
deformations of inhomogeneous groups via projective space approach
was considered in \cite{Sch}. In this approach inhomogeneous
transformations are represented by block-triangular matrix
so that general theory of matrix quantum groups can be applied.
Unfortunately, the standard quantum deformation in this case leads to the
appearance of additional dilatations.

Another way to attack the problem is to start from quantum deformation
of Poincar\'e Lie algebra \cite{OWZ2}-\cite{KobU}. As is well known
we can consider q-Lie algebra from two closely related point of view
\cite{FRT}: as quantum universal enveloping (QUE) algebra and as dual
space to q-group space.
According to these forms there are two approaches to q-Poincar\'e
algebra problem.

In one of them \cite{OWZ2} q-Poincar\'e
algebras are derived from the action of generators on the
noncommuting coordinates of quantum space-time. One parametric
deformation once more forces to include dilatations (note that the
authors of \cite{KobU} do not meet this problem because they do not
consider coproduct and hence the full Hopf algebra structure of
q-Poincar\'e algebra).

Another approach \cite{Luk} is based on contraction of QUE
anti-de Sitter algebra $U_q(O(3,2))$, the deforming parameter $q$ being
also contracted \cite{Fir}. As a result one obtains QUE Poincar\'e
algebra and
deformed dynamics of quantum field theoretical model (deformed
Klein-Gordon equation) defined on the space-time with
usual commutative geometry. Important problem here is to find
corresponding q-group of space-time transformations (contraction on
q-group level is not well understood yet, especially in the case of
q-parameter contraction). The attempts to find QUE Poincar\'e
algebra as a Hopf subalgebra of $U_q(O(4,2))$ or $U_q(SU(2,2))$
\cite{Dob},\cite{Luk2} meet the same problem as in the case of
inhomogeneous q-group: one has to add generators of dilatations
to define selfconsistent coproduct. In \cite{Dob} the general
consideration of the problem based on Cartan automorphisms and Bruhat
decomposition of non-compact algebras was done in the context of canonical
procedure for q-deformation of non-compact Lie algebras. In
particular, it was shown that Poincar\'e containing Hopf subalgebra
of the quantum conformal algebra has to include dilatations, i.e. to
be the 11-generators Weyl algebra (parabolic-type subalgebra).

We would like to note that it is very natural to look for
q-Poincar\'e group as a q-subgroup of q-conformal one because of at
least two reasons: {\it i}) to start from $SU(2,2)$ is very desirable
from the physical point of view because of well known important
role of the conformal group at high energy and small distances where
we hope noncommutative geometry becomes essential (direct construction
of the q-Poincar\'e group does not provide automatically that it is
a q-subgroup of some quantum deformation of conformal group);
{\it ii}) conformally invariant field theory has the problems because
of ultra-violet divergences, but the most attractive aim of
q-deformation is to remove any divergences, so to start from
q-conformal symmetry is logically self-consistent approach.

In this paper we apply general idea of multiparametric deformation
for obtaining the inhomogeneous groups developed in our previous work
\cite{CD}, to physically interesting case of the q-Poincar\'e group.
To avoid q-group contractions, we use as a starting point the twistor
approach (see, e.g. \cite{Pen}) to the definition of Minkowski space and
Poincar\'e group.

\section{Quantum deformation of the Poincar\'e group}

\indent Let us remind some basic facts concerning ${\bf G(2,4)}$ and its
transformations by conformal and Poincar\'e group in classical case
\cite{Pen}. The Grassmann manifold  ${\bf G(2,4)}$ is
a set of 2-dimensional complex subspaces ${\bf C^2}$ of 4-dimensional
space ${\bf C^4}$. The homogeneous coordinates of ${\bf G(2,4)}$
form $2 \otimes 4$ matrix $Z$ with matrix elements ${Z_{\ \alpha}^j}
(\alpha = 1,2; j=1,...,4)$.
Grassmann manifold is topologically non-trivial and can be covered
by six patches (big cells) $U_{ij} (i<j;\ i,j=1,...,4)$ defined by the
conditions
$$Z \in U_{ij} \Longleftrightarrow {\det} Z_{(ij)} \not= 0.$$
The $2 \otimes 2$ submatrix $Z_{(ij)}$ is formed by $i$ and $j$ rows of
matrix $Z$. On any big cell $U_{ij}$ one can introduce
inhomogeneous coordinates
$z=Z_{(kl)}Z^{-1}_{(ij)}$ (here $k,l \not= i,j$). In fact each
$U_{ij}$ is a complexified Minkowski space  with coordinates $x^\mu
\in \bf{C}$ defined by the relation $z=x^\mu {\sigma}_\mu$
(${\sigma}_\mu$ are the Pauli matrices). If $z=z^+$, then $x^\mu \in {\bf
R}$ are the coordinates of the usual Minkowski space ${\cal M}$.
The action of the Poincar\'e group is deduced from the action of
$SU(2,2)$ on $G(2,4)$:
\begin{equation}
Z \rightarrow Z^\prime = TZ, \qquad T \in SU(2,2).   \label{2.1}
\end{equation}

Let us choose the big cell $U_{34}$ for definiteness, write the matrix $T \in
SU(2,2)$ in a block form,
\begin{equation}
T
= \left( \begin{array}{cc}
A & B \\
C & D
\end{array} \right)\ ,                               \label{2.2}
\end{equation}
where $A,B,C,D$ are $2 \otimes 2$ matrices and choose hermitian
metric $\Phi$ of the group $SU(2,2)$ in the form
\begin{equation}
\Phi
= \left( \begin{array}{cc}
0 & i{\bf 1}_2 \\
-i{\bf 1}_2 & 0
\end{array} \right)  .                                \label{2.3}
\end{equation}
Matrices $T$ obey the unitarity condition
\begin{equation}
    T^+\Phi T = \Phi\ .                                  \label{2.4}
\end{equation}
The Poincar\'e subgroup of the $SU(2,2)$ is defined by
\begin{eqnarray}
  C = 0\ , \nonumber\\
  {\det} A = 1\ .                                          \label{2.6}
\end{eqnarray}
Note that such simple and convenient form of conditions
distinguishing Poincar\'e group out of conformal one is possible due
to antidiagonal form of the metric $\Phi$.
In this case we have from (\ref{2.4}) and (\ref{2.6})\ ,
\begin{equation}
  B^+D = D^+B\ ,                                     \label{2.7}
\end{equation}
\begin{equation}
  A^+ = D^{-1}\ ,                                     \label{2.8}
\end{equation}
\begin{equation}
  {\det} A = {\det} D =1\ .                                     \label{2.9}
\end{equation}
The group action (\ref{2.1}) on the inhomogeneous coordinates $z$
takes the form
\begin{equation}
z \rightarrow z^{\ \prime} = AzA^+ + BA^+ .           \label{2.10}
\end{equation}

The hermitian matrices $z \in {\cal M}$ are transformed by (\ref{2.10}) to the
hermitian ones because of conditions (\ref{2.7}),(\ref{2.8}) and the Minkowski
length $ds^2= {\det} (z_1 -z_2)$ is invariant. Thus (\ref{2.10}) are
completely equivalent to the usual Poincar\'e group transformations.

To "quantize" this construction we have to find, first of all,
a real form $SU(2,2)_{q_{ij}}$ of
$SL(4,C)_{q_{ij},r}$, corresponding to
the hermitian metric
(\ref{2.3}). In other words, we must find a multiparametric R-matrix
consistent with involution,
\begin{equation}
T^* = \Phi (T^{-1})^t \Phi ,                         \label{2.11}
\end{equation}
followed from unitarity condition (\ref{2.4}). Here $(T^{-1})_{ij}$ is
antipode
of $T_{ij}$ defined in a usual way for $SL(4,C)_{q_{ij},r}$ group \cite{FRT},
\cite{Schirr}.
 Multiparametric R-matrix for $GL(4,C)_q$ has a
form \cite{Schirr}
\begin{equation}
R^{ij}_{\quad kl} = \delta^i_{\ k}\delta^j_{\ l}\bigl(\delta^{ij} +
\Theta^{ji} q^{-1}_{ij} + \Theta^{ij} q_{ji}r^{-2} \bigr) +
\delta^i_{\ l}\delta^j_{\ k}\Theta^{ij}\bigl(1-r^{-2}\bigr) ,
                                                      \label{2.12}
\end{equation}
$\Theta^{ij} = 1$ for $i>j$ and 0 else.
As usual, for the $SU(m,n)$ case we put $q_{ij}, r^2 \in
{\bf R}$ \cite{FRT}, \cite{Schirr2} and thus from defining relations
for $T_{ij}$,
\begin{equation}
R_{ik,rs}T_{rv}T_{sw} = T_{kb}T_{ia}R_{ab,vw}\ ,         \label{2.13}
\end{equation}
we obtain equations
$$R_{ik,rs}T^*_{sw}T^*_{rv} = T^*_{ia}T^*_{kb}R_{ab,vw} ,$$
which can be written due to (\ref{2.11}) in the form
\begin{equation}
\tilde{R}_{sr,ki}T_{rv}T_{sw} = T_{kb}T_{ia}\tilde{R}_{wv,ab} ,
                                                       \label{2.14}
\end{equation}
where
\begin{equation}
\tilde{R}_{sr,ki} = \Phi _{sa}\Phi _{rb}R_{ab,cd}\Phi _{ck}
\Phi _{di} .                                           \label{2.15}
\end{equation}
Comparison of (\ref{2.13}) and (\ref{2.14}) yields the identity
$$ \tilde{R}_{sr,ki} = R_{ik,rs}\ , $$
which in turn gives for deformation parameters the conditions
$$r^2 =1 ,\  q_{ij} = q_{\hat{j}\hat{i}} \qquad
(\hat{i}=5-i,\hat{j}=5-j) , $$
followed from the explicit form of the R-matrix (\ref{2.12}).
In our case of $GL(4,C)_{q_{ij},r}$ group, this means
\begin{equation}
 r^2 =1 ,\ q_{12} = q_{34} ,\ q_{13} = q_{24}\
(q_{ij} \in {\bf R}).                                  \label{2.16}
\end{equation}

We would like to stress that $r^2=1$ condition is the consequence of
the antidiagonal form of the metric $\Phi$, which is
appropriate for picking out the Poincar\'e subgroup.
As we shall show later, the triviality of parameter $r$
leads to extremely simple structure of the corresponding deformed
Lie algebra. This is in accordance with Drinfeld's theorem
\cite{Drinf} on uniqueness of the Lie-algebra deformation (see also
\cite{SchirrWZ}).

To construct q-Poincar\'e subgroup ${\cal P}_q$ of SU(2,2) we have
to require centrality of ${\det}_q A$ and ${\det}_q D = {\det}_q (A^+)^{-1}$
according to condition (\ref{2.9}). As is shown in \cite{CD2} this
leads to the conditions
\begin{equation}
q_{12} = q_{34} = 1,\quad  q_{14} = q_{23} \equiv q,\quad
q_{13} = q_{24} \equiv q^{-1}\ ,                           \label{2.19}
\end{equation}
which does not contradict the real form conditions (\ref{2.16}).
Taking the deformation parameters according to
(\ref{2.19}), we can put
${\det} A = {\det} D = 1$ to obtain consistent q-Poincar\'e subgroup
${\cal P}_q$ of $SU(2,2)_q$ with a structure of real Hopf algebra
inherited from $SU(2,2)_q$ \cite{CD2}.

We can further restrict our q-group to obtain Lorentz subgroup formed
by the matrices
\begin{equation}
T
= \left( \begin{array}{cc}
A & 0 \\
0 & (A^+)^{-1}
\end{array} \right)\ .                               \label{2.24}
\end{equation}
Matrix elements of $A,\ A^+$ belong to *-Hopf subalgebra of ${\cal
P}_q$, so matrices (\ref{2.24}) define q-Lorentz subgroup of ${\cal
P}_q$. Very unusual property of our deformation is the commutativity of
elements of $A$ with each other and $A^+$ with each other. The only
nontrivial CR are those between elements of $A$
and $A^+$. This is the simplest possible deformation of Lorentz group.

Condition $A=(A^+)^{-1} $
 picks out $SO(3)_q$ rotation subgroup. But our deformation is
inconsistent with such condition \cite{CD2} because elements of
$A$ between themselves and with those of $A^+$ commute differently.
So there is no q-rotation subgroup of q-Lorentz group in our case.
The absence
of this subgroup is not very important from the physical
point of view because, as we hope, q-Poincar\'e symmetry has meaning
at extremely high energy, so that $SO(3)_q$ subgroup can not play
essential physical role. Of course, this fact can have strong
influence on q-Poincar\'e representation theory since subgroup of
rotations is a small group of fixed momentum for massive states in
$q=1$ case and so is important ingredient for induced representations of
the Poincar\'e group.

\section{Quantum Minkowski geometry}

\indent The commutation relations for homogeneous coordinates $Z$ are
the same as for the two last columns of the matrix $T$. In this case
they are saved under the group
transformations. Moreover, due to the centrality of ${\det} Z_{(34)}$ (in
analogy with ${\det} D$) we can put ${\det} Z_{(34)} =1$ and define
inhomogeneous coordinates
$$ z = Z_{(12)}Z^{-1}_{(34)} \equiv
\left ( \begin{array}{cc}
z^1 & z^4 \\
z^2 & z^3
\end{array} \right )\ . $$
Using once more the CR (\ref{2.13}),
we obtain
\begin{equation}
z^iz^j=Q_{ij}z^jz^i\ ,                      \label{3.24}
\end{equation}
where
$$
Q_{ij}=q^2 \quad \mbox{if} \quad j=i+1(mod4),\qquad
Q_{ji}=Q_{ij}^{-1}\ ,                                   $$

$$
Q_{ij}=1,   \quad \mbox{if} \quad j=i+2(mod4)\ \mbox{or} \ i=j\ .
$$

To obtain deformed Minkowski length $l_q$ which is invariant
under q-Poincar\'e transformations, we have to take into account
that the matrix elements of submatrix $A$ do not commute with
those of $A^+$ so we can not put $l_q = {\det}\ z$ though
this is a central element of the algebra, but assume a more
general expression
$l_q = {\det}_{q^n} z \equiv z_1z_4 - q^nz_2z_3\ ,$
with some integer $n$ ($l_q$ is central for any $n$).
Straightforward calculations show that if $n=2$ then the determinant
factorizes
$$ {\det}_{q^2} (AzA^+) = ({\det} A) ({\det}_{q^2} z)({\det}\ A^+)
= {\det}_{q^2} z
$$
and the deformed Minkowski length
$$l_q = {\det}_{q^2} z = g^q_{ij}z^iz^j$$
is invariant under homogeneous transformations due to the
conditions ${\det} A = {\det} A^+=1$. Here $g^q_{ij}$ is q-analog
of classical $SO(2,2)$ metric $g_{ij}$:
\begin{equation}
g^q_{ij}=
\left( \begin{array}{cccc}
0 & 0 & 1 & 0 \\
0 & 0 & 0 & -q^2 \\
1 & 0 & 0 & 0 \\
0 & -q^2 & 0 & 0
\end{array} \right)\ .                                    \label{3.24aa}
\end{equation}
Note that we consider the quantum analog of usual Minkowski space but
for the sake of convenience use coordinates with other reality conditions
\begin{equation}
(z^1)^* = z^1 ,\quad  (z^3)^* = z^3 ,\quad
(z^2)^* = z^4 ,\quad (z^4)^* = z^2 ,               \label{3.24b}
\end{equation}
in contrast to the usual pure real ones $(x^{\mu})^*=x^{\mu}$.
One can straightforwardly check that the involution (\ref{3.24b}) is
consistent with the defining relations (\ref{3.24}).
The inhomogeneous transformations have the form
\begin{equation}
z \rightarrow ({z^\prime})^\alpha_{\ \beta} = (AzA^+)^\alpha_{\
\beta}  + (BA^+)^\alpha_{\ \beta}                       \label{3.24c}
\end{equation}
and direct inspection gives
\begin{equation}
[(AzA^+)^\alpha_{\ \beta} , (BA^+)^\alpha_{\ \beta}] = 0\quad
\forall \alpha,\beta=1,2 .                                \label{3.25}
\end{equation}
{}From this relation we can deduce the affine structure of the quantum
Minkowski space: identity (\ref{3.25}) implies that we can
consider a set of coordinates $(z_t)^\alpha_{\ \beta}$ , where $t$
is discrete or continuous index to distinguish different "points" of
q-space-time and put
\begin{equation}
z^i_tz^j_{t^\prime}=Q_{ij}z^j_{t^\prime}z^i_t \qquad \forall\
t,\ t^\prime\ .
\end{equation}
Now we can define the invariant of the complete Poincar\'e group:
\begin{equation}
ds^2_q = {\det}_{q^2} (z_t - z_{t^\prime}) .                 \label{3.26}
\end{equation}
Coordinates of q-Minkowski space ${\cal M}_q$ form the vector
representation of ${\cal P}_q$. Construction of complete theory
of  ${\cal P}_q$-representations
is a complicated problem because of necessity to develop integral
calculus on noncommutative space and due to the absence of rotational subgroup
(the small subgroup for massive states in classical case). But
spinorial representations of q-Lorentz subgroup can be derived rather
easily in analogy with the usual ones (see e.g. \cite{Spin}).

A spinor $\xi_{\alpha}$ and its conjugate $\bar \xi^{\dot \alpha}$
($\alpha=1,2$) are transformed by q-Lorentz submatrix
according to
$$
\left( \begin{array}{c}
\xi^\prime \\
{\bar \xi}^\prime
\end{array} \right) =
\left( \begin{array}{cc}
A & 0 \\
0 & (A^+)^{-1}
\end{array} \right)\
\left( \begin{array}{c}
\xi \\
{\bar \xi}
\end{array} \right)\ .
$$
As we have already mentioned matrix elements of $A$ commute with each
other. Hence $\xi_1\xi_2=\xi_2\xi_1$. The same is true for
$(A^+)^{-1}$ and components of $\bar \xi$. So q-Lorentz invariants
$\xi_{\alpha}\xi_{\beta}\epsilon^{\alpha\beta}$ and
$\bar\xi^{\dot\alpha}\bar\xi^{\dot\beta}\epsilon_{\dot\alpha\dot\beta}$
can be constructed with help of usual antisymmetric
tensors $\epsilon_{\alpha\beta}$ and $\epsilon_{\dot\alpha
\dot\beta}$. Thus lowest spinors do not differ from the undeformed ones.
The q-relations between $\xi$ and $\bar\xi$ are the same as those for
the matrix elements of any column of matrix $Z$ or $T$
\begin{equation}
\xi_\alpha\bar\xi^{\dot\beta}=q_{\alpha (\beta +2)}
\bar\xi^{\dot\beta}\xi_\alpha\ ,                     \label{3.26aa}
\end{equation}
or, using (\ref{2.19}),
$$
\begin{array}{ll}
\xi_1\bar\xi^{\dot 1}=q^{-1} \bar\xi^{\dot 1}\xi_1\ , &
\xi_1\bar\xi^{\dot 2}=q \bar\xi^{\dot 2}\xi_1\ , \\
\xi_2\bar\xi^{\dot 1}=q \bar\xi^{\dot 1}\xi_2\ , &
\xi_2\bar\xi^{\dot 2}=q^{-1} \bar\xi^{\dot 2}\xi_2\ .
\end{array}
$$
We can generalize the relations (\ref{3.26aa}) as in the case of
coordinates and introduce the relations for different spinors
$$
\xi_\alpha\bar\eta^{\dot\beta}=q_{\alpha (\beta +2)}
\bar\eta^{\dot\beta}\xi_\alpha\ ,                      $$
$$
\xi_\alpha\eta_{\beta}=\eta_{\beta}\xi_\alpha\ ,  \qquad
\bar\xi^{\dot\alpha}\bar\eta^{\dot\beta}=
\bar\eta^{\dot\beta}\bar\xi^{\dot\alpha}\ .            $$
Now the higher spin representations can be constructed by a
method close to that in the classical case \cite{Spin}.
Let us consider non-normalized symmetrical spinors
\begin{equation}
F_{\sigma
\dot\rho}(j_1,j_2)=
\xi_1^{j_1+\sigma}\xi_2^{j_1-\sigma}\bar\eta_{\dot
1}^{j_2+\dot\rho}\bar\eta_{\dot 2}^{j_2-\dot\rho} \ ,     \label{3.26a}
\end{equation}
where $-\infty<j_1,j_2<\infty,\ \sigma\leq\mid j_1\mid ,\
\rho\leq\mid j_2\mid $. They form representation space $V_{j_1,j_2}$
of q-Lorentz group of dimension $(2j_1+1)(2j_2+1)$. Transformations
of this spinors $F(j_1,j_2)$ are defined by transformations of
spinors $\xi$ and $\bar\eta$
$$
F^\prime _{\sigma
\dot\rho}(j_1,j_2)=
(A\xi)_1^{j_1+\sigma}(A\xi)_2^{j_1-\sigma}(A^*\bar\eta)_{\dot
1}^{j_2+\dot\rho}(A^*\bar\eta)_{\dot 2}^{j_2-\dot\rho} \equiv
{\cal D}(A)_\sigma^{\ \lambda}F_{\lambda\dot\tau}(j_1,j_2)
{\cal D}((A^+)^{-1})^{\dot\tau}_{\ \dot\rho}\ ,
$$
the representation matrices ${\cal D}(A),\ {\cal D}((A^+))$ being the
same as in the classical case due to commutativity of elements of
matrix $A$ with each other.
This is similar to the homogeneous part of coordinates
transformations (\ref{3.24c}). Note, however, that to obtain
non-isotropic Minkowski vector one has to use two different spinors
(two-columns matrix $Z$) and their conjugates \cite{Pen}. So
coordinates $z^i$ correspond to non-symmetrical spinors. Quantization
leads to the deformation of invariant metric on the space
$V_{j_1j_2}$ in the same way as on the q-Minkowski space.
Consider the spinors with upper indices
$$
\Phi^{\sigma
\dot\rho}(j_1,j_2)=
(\zeta^1)^{j_1+\sigma}(\zeta^2)^{j_1-\sigma}(\bar\theta^{\dot
1})^{j_2+\dot\rho}(\bar\theta^{\dot 2})^{j_2-\dot\rho} \ ,
$$
Taking into account the CR between spinors and
conjugated spinors one can show, that the invariant metric has the form
$$
<F,\Phi>=\sum_{\sigma,\ \dot\rho} a^q_{\sigma\dot\rho}(j_1,j_2)
F_{\sigma\dot\rho}(j_1,j_2)\Phi^{\sigma\dot\rho}(j_1,j_2)\ ,
$$
where
$$
a_q(j_1,j_2;\sigma,\dot\rho)={{(2j_1)!(2j_2)!}\over
{(j_1+\sigma)!(j_1-\sigma)!(j_2+\rho)!(j_2-\rho)!}}q^{-4\sigma\rho}\ .
$$
In the classical limit $q=1$ one obtains usual normalization factor
$\sqrt{a(j_1,j_2)}$ of symmetrical spinors.

Now we can obtain realization of q-Poincar\'e
transformations on functions $\psi(z)$ of noncommuting coordinates
(i.e. fields from the physical point of view). Let
$\psi_{\sigma\dot\rho} \in V_{j_1j_2}$ and so is transformed by
q-Lorentz group as multispinor (\ref{3.26a}). Then q-Poincar\'e
transformations can be realized on the maps $\psi : {\cal M}_q
\longrightarrow  V_{j_1j_2}$ in the way quite
analogous to the classical case (cf. e.g. \cite{BD}):
\begin{equation}
\psi^\prime_{\alpha\dot\rho}(z)={\cal D}(A)_\sigma^{\ \lambda}
\psi_{\lambda\dot\tau}(z^{\prime\prime})
{\cal D}((A^+)^{-1})^{\dot\tau}_{\ \dot\rho}\ ,      \label{3.26b}
\end{equation}
where $z^{\prime\prime}$ is the coordinate matrix transformed by
antipode $A^{-1}$
$$
z^{\prime\prime}=A^{-1}z(A^+)^{-1}-A^{-1}B\ .       $$

\section{q-Poincar\'e algebra}

\indent As generators of q-translations it is natural to use q-derivatives
(cf. \cite{OWZ2}). Thus first of all, we have to derive formulae for
differential calculus on our Minkowski space. This can be done in
analogy with \cite{WZ},\cite{OWZ2} and gives the result
\begin{equation}
\partial_i\partial_j=Q_{ij}\partial_j\partial_i\ ,     \label{4.0}
\end{equation}
$$
dz^idz^j=-Q_{ij}dz^jdz^i\ ,  $$
$$
\partial_jz^i=\delta^i_j+Q_{ij}z^i\partial_j\ ,      $$

To find q-Poincar\'e algebra, we can start from general anzatz for the
action of q-Lorentz generators $L$ on Minkowski coordinates,
\begin{equation}
Lz^i=\alpha_iz^iL+A^i_jz^j\ .                   \label{4.1a}
\end{equation}
The first term in the r.h.s. is diagonal because of the diagonal R-matrix.
The unknown matrix elements $A^i_j$ and $\alpha_i$ can be defined from
conditions of metric invariance,
\begin{equation}
Ll_q=l_qL                                 \label{4.2}
\end{equation}
and from the defining relations invariance,
\begin{equation}
L(z^iz^j-Q_{ij}z^jz^i)=(z^iz^j-Q_{ij}z^jz^i)L\ .  \label{4.3}
\end{equation}

One can show that these conditions leads to unique solution for
q-generators (\ref{4.1a}) (i.e. unique solution for $\alpha_i$ and
$A^i_j$).
Obviously, it would be desirable to find a form of q-Lie algebra CR
analogous to the covariant
tensorial form of nondeformed Lorentz Lie algebra ($SO(2,2)$ in our case):
\begin{equation}
[M^{mn},M^{pk}]=g^{mk}M^{np}+g^{np}M^{mk}-g^{mp}M^{nk}-g^{nk}M^{mp}\ ,
\label{4.6}
\end{equation}
where $g^{mk}$ is inverse to (\ref{3.24aa}) at $q=1$. This can
be achieved most easily by constructing the spinless representation of
q-Lie algebra by q-differential operators, i.e. by the deformation
of usual Killing vectors
$$M^{mn}=z^m\partial^n-z^n\partial^m,\qquad
\partial^m=g^{mn}\partial_n \ ,$$
on commuting Minkowski space.
For this aim we introduce the two-index q-Lorentz generators
$L^{mn}=-Q_{mn}L^{nm}$
and put
\begin{equation}
L^{mn}=z^m\partial ^n-Q_{mn}z^n\partial ^m,\qquad
\partial ^m=g^{mn}_q \partial _n \ .                    \label{4.7}
\end{equation}
One can check that (\ref{4.7}) satisfies (\ref{4.2}),(\ref{4.3})
and is in one to one correspondence with the unique solution for
the anzatz (\ref{4.1a}).
For the CR of $L^{mn}$ one can
deduce the desirable form:
\begin{equation}
[L^{mn},L^{pk}]_{Q(mn,pk)}=g^{mk}_qQ_{mn}Q_{pk}L^{np}+g^{np}_qL^{mk}-
g^{mp}_qQ_{mn}L^{nk}-g^{nk}_qQ_{pk}L^{mp}\ ,
                                               \label{4.8}
\end{equation}
where we used the symbol
$$
Q(ijk...,lmn...)\equiv Q_{il}Q_{im}Q_{in}...Q_{jl}Q_{jm}
Q_{jn}...Q_{kl}Q_{km}Q_{kn}...\ .
$$
Here factors in r.h.s contain all ordered pairs of indices from two
sets in l.h.s divided by comma. In particular,
$Q_{mn,pk}\equiv Q_{mp}Q_{mk}Q_{np}Q_{nk}$.

Now we can add the generators of q-translations, i.e. q-derivatives, to
complete q-Poincar\'e Lie algebra:
\begin{equation}
[\partial^k,L^{mn}]_{Q(k,mn)}=g^{km}_q\partial^n-
g^{km}_qQ_{mn}\partial^m\ .       \label{4.9}
\end{equation}
Coproduct $\Delta$ for generators of q-Poincar\'e algebra can be read
off from the action on monoms
$m(A,B,C,D)\equiv (z^1)^A(z^3)^B(z^2)^C(z^4)^D$ ($A,B,C,D$ are
arbitrary integer numbers). For example,
$\Delta (L^{23})=L^{23}\otimes {\bf 1}+q^{2(q^2L^{24}-L^{13})}\otimes
L^{23};\ \
\Delta (\partial_1)=\partial_1\otimes {\bf 1}+q^{2q^2L^{24}}\otimes
\partial_1$.
Counity $\epsilon$ is defined, as usual,
$\epsilon(L^{mn})=\epsilon(\partial _k)=0$,
and antipode S is defined from the defining property $\mu (id\otimes
S)\Delta(L^{mn})= \mu (S\otimes id)\Delta(L^{mn})=0$
(here $\mu$ is the algebra multiplication).

As we noted earlier, according to the Drinfeld's uniqueness theorem
\cite{Drinf} our q-Lie algebra has to be simply connected with usual,
nondeformed one because Lie algebra deformation parameter is trivial.
{}From the geometrical and physical point of view the
most natural way to establish such correspondence is the
following.

Let us introduce q-tetrade in the q-space-time ${\cal M}_q$, i.e. four
q-vectors $e_a^{\ n}$, so that
\begin{equation}
e_a^{\ n}e_b^{\ m}=Q_{nm}e_b^{\ m}e_a^{\ n} \qquad  (a,b=1,...,4).
                                                        \label{4.12}
\end{equation}
One can show that q-determinant defined by
\begin{equation}
e\equiv \epsilon^q_{mnpk}e_1^{\ m}e_2^{\ n}e_3^{\ p}e_4^{\ k}=
\epsilon^{abcd}e_a^{\ 1}e_b^{\ 2}e_c^{\ 3}e_d^{\ 4}
                                                         \label{4.13}
\end{equation}
and is q-Poincar\'e invariant and central element of the algebra.
Here $\epsilon^q_{ijkl}$ is q-deformed antisymmetric Levi-Cevita
tensor: $\epsilon^q_{1234}=1$ and
$\epsilon^q_{ijkl}=-Q_{ji}\epsilon^q_{jikl}=
-Q_{kj}\epsilon^q_{ikjl}=-Q_{lk}\epsilon^q_{ijlk}\ .       $
Now it is possible to introduce commuting coordinates $z^a$
on ${\cal M}_q$ through the relations
\begin{equation}
z^m=z^ae_a^{\ m}
                                                         \label{4.15}
\end{equation}
or
\begin{equation}
z^a=e^a_{\ m}z^m\ ,
                                                        \label{4.16}
\end{equation}
where $e^a_{\ m}$ are the elements of the inverse matrix of the
q-tetrade defined with help of q-determinant $e$
\begin{equation}
e^a_{\ m}=(e^{-1})\epsilon^q_{l...rmp...k}
e_1^{\ l}...e_{a-1}^{\ r}e_{a+1}^{\ p}...e_4^{\ k}Q(l...r,m)
                                                        \label{4.17}
\end{equation}
(generalization of the inverse matrix formula in usual case).
It is easy to see that
\begin{equation}
e_m^{\ a}e_n^{\ b}=Q_{mn}e_n^{\ b}e_m^{\ a}
                                                        \label{4.18}
\end{equation}
and that $z^a\ (a=1,...,4)$ are commuting coordinates. Such
q-tetrade and commuting coordinates can be introduced in the q-space
with commutative affine structure only, i.e. if vectors of
translation commute with homogeneous part of coordinates (see
(\ref{3.25})). We expect that in general q-affine space CR of $z^a$
are defined by parameter $r$ and it may be useful to separate different
parameters of deformation with help of analogous q-bein.

Now we are ready to show why the existence of globally defined
q-tetrade leads to trivial q-algebra but non-trivial q-group. Indeed,
using differential operator representation (\ref{4.7}) one has
\begin{equation}
L^{mn}=e_a^{\ m}z^ae_b^{\ n}\partial^b-
Q_{mn}e_b^{\ n}z^be_a^{\ m}\partial^a=
e_a^{\ m}e_b^{\ n}(z^a\partial^b-z^b\partial^a)=
e_a^{\ m}e_b^{\ n}M^{ab}
                                         \label{4.19}
\end{equation}
where we used
$$
\partial^m=e_a^{\ m}\partial^a\ .           $$
Obviously one can define inverse transformation
$$
M^{ab}=e_n^{\ b}e_m^{\ a}L^{mn}                  $$
Now all the properties of the q-algebra including commutators,
Casimir operators, etc. can be derived from the well known properties
of the usual Poincar\'e algebra of $(M^{ab},\partial^c)$.

{}From the other hand the relation is not so simple in the group
transformations case. For the sake of simplicity let us consider
homogeneous transformations only (inclusion of the translations does
not change anything). In vector notations the q-Lorentz
transformations has a form
\begin{equation}
z^m=(\Lambda_q)^m_{\ n}\otimes z^n\ .
                                          \label{4.20}
\end{equation}
Substituting  (\ref{4.15}) in (\ref{4.20}) one obtains
$$
z^a=\Lambda^a_{\ b}z^b                   $$
with
$$
\Lambda^a_{\ b}=e^b_{\ m}(\Lambda_q)^m_{\ n}\otimes e^n_{\ a}  $$
and the matrix elements of $\Lambda^a_{\ b}$ commute with each other
if $$
e^b_{\ n}(\Lambda_q)^m_{\ k}=Q_{mn}(\Lambda_q)^m_{\ k}e^b_{\ n}\ .
$$
However, it is impossible to derive inverse formula,
i.e. to construct $\Lambda_q$ from $\Lambda$. Indeed,
$$
z^{\prime m}=(\Lambda_q)^m_{\ n}\otimes z^n=e^m_{\ a}z^{\prime a}=\\
(e^m_{\ a}\Lambda^a_{\ b}e^b_{\ n})z^n\equiv S^m_{\ n}z^n\ .
$$
But $S^m_{\ n}$ nontrivially commute with $z^k$ and, hence, can not be the
element of the quantum group transformation matrix $\Lambda_q$ by
definition. This shows non-trivial character of the q-Poincar\'e group
deformation. Another important property of the q-deformation that
confirms this statement is the absence of the $O(3)_q$-subgroup.

\section{Conclusion}

\indent Starting from the quantum conformal group $SU(2,2)_q$,
we have defined a self-consistent quantum Poincar\'e group
as its q-subgroup, q-Minkowski space and Lie algebra. We present the
form of higher spinor q-Lorentz representations, realization of
q-Poincar\'e transformations on the fields defined on q-Minkowski
space. The essential points of our construction are twistor
approach to the definition of Minkowski space and multiparametric
deformation of the linear projective groups.
We would like to stress that if one considers q-Poincar\'e group as a
q-subgroup of quantum conformal group and use known multiparametric
R-matrix \cite{Schirr}, one should obtain the group we
have constructed. It has diagonal R-matrix and commuting affine
structure of corresponding q-Minkowski space. This leads to great
simplification of most formulae and make them quite analogous to the
classical ones. Due to commuting affine structure one can introduce
globally defined q-tetrade that connects commuting and q-commuting
coordinates as well as usual and q-Lie algebra.
We hope that such relatively simple form of
deformation will be useful for construction of induced representation
theory of the deformed Poincar\'e group, developing the q-deformed
field theory etc.
As an illustration of the application of our q-group to the general
analysis of the q-deformed quantum field theory let us consider very
briefly the relation between q-deformed space-time symmetry and
q-oscillators. Remind that usual quantum fields $\psi(z)$ have the
following relation with creation and destruction operators
$a^\dagger ({\bf p}),\ a({\bf p})$ (we consider the real field for
simplicity)
\begin{equation}
\psi^N(z)=
(2\pi )^{-3/2}\int \left[
v^{\dagger N}_{\ A}({\bf p})a^A({\bf p})e^{-i(pz)}+
v^ N_{\ B}({\bf p})a^{\dagger B}({\bf p})e^{i(pz)}\right]
\frac{d{\bf p}}{2\omega({\bf p})}
                                        \label{5.1}
\end{equation}
(in our simple deformation case one can define integration due to
relation with commuting variables).
Here $N$ is (multi)index of the representation and $v^N_{\ A}$ is the
corresponding wave function (see, e.g.\cite{BD}). One can see easily
the analogy between (\ref{5.1}) and the relation (\ref{4.15}) for
q-coordinates $z^m$ and commuting one $z^a$, the wave function
$v^N_{\ A}$ playing the role of the q-bein in the space of the
representation. So in our case creation and destruction operators
obey the usual CR in analogy with commuting coordinates $z^a$.
Thus in general case non-trivial q-oscillator relations can be
connected with algebra deformation parameter $r$ only, and CR between
components of wave functions are defined by other (existing in our
case) deformation parameters.

\end{document}